\documentclass[twocolumn]{aastex63}
\usepackage{amsmath}
\usepackage{multirow}
\usepackage[normalem]{ulem}

\newcommand{\nicer}{{\it NICER}}
\newcommand{\Msun}{\,{\rm M}_{\odot}}

\begin{document} 
\title{Constraints on the maximum mass of neutron stars with a quark core from GW170817 and {\it NICER} PSR J0030+0451 data}

\correspondingauthor{Ang Li}
\email{liang@xmu.edu.cn}

\author{Ang Li}
\affiliation{Department of Astronomy, Xiamen University, Xiamen, Fujian 361005, China}

\author{Zhiqiang Miao}
\affiliation{Department of Astronomy, Xiamen University, Xiamen, Fujian 361005, China}

\author{Sophia Han}
\affiliation{Institute for Nuclear Theory, University of Washington, Seattle, WA~98195, USA}
\affiliation{Department of Physics, University of California Berkeley, Berkeley, CA~94720, USA}

\author{Bing Zhang}
\affiliation{Department of Physics and Astronomy, University of Nevada Las Vegas, Las Vegas, NV 89154, USA}

\date{\today}

\begin{abstract} 
We perform a Bayesian analysis of the maximum mass $M_{\rm TOV}$ of neutron stars with a quark core, incorporating the observational data from tidal deformability of the GW170817 binary neutron star merger as detected by LIGO/Virgo and the mass and radius of PSR J0030+0451 as detected by \nicer.
The analysis is performed under the assumption that the hadron-quark phase transition is of first order, where the low-density hadronic matter described in a unified manner
by the soft QMF or the stiff DD2 equation of state (EOS)  
transforms into a high-density phase of quark matter modeled by the generic ``Constant-sound-speed'' 
(CSS) parameterization.
The mass distribution measured for the $2.14 \Msun$ pulsar, MSP J0740+6620, is used as the lower limit on $M_{\rm TOV}$.
We find the most probable values of the hybrid star maximum mass are $M_{\rm TOV}=2.36^{+0.49}_{-0.26}\Msun$ ($2.39^{+0.47}_{-0.28}\Msun$) for QMF (DD2), 
with an absolute upper bound around $2.85\Msun$, to the $90\%$ posterior credible level. Such results appear robust with respect to the uncertainties in the hadronic EOS.
We also discuss astrophysical implications of this result, especially on the post-merger product of GW170817, short gamma-ray bursts, and other likely binary neutron star mergers.
\end{abstract}


\section{Introduction}
\label{sec:intro}

Recently, an increasing number of binary systems are found to have at least one compact object falling into the possible gap between neutron star (NS) and black hole (BH) masses~\citep{2010ApJ...725.1918O}, e.g. the compact remnants of two NS-NS merger events GW170817~\citep{2017PhRvL.119p1101A} ($\sim2.73\Msun$) and GW190425 ($\sim3.4\Msun$)~\citep{2020ApJ...892L...3A}, as well as the secondary component 
in the binary merger event GW190814 ($\sim2.6\Msun$)~\citep{2020ApJ...896L..44A}.
There are also recent reports on the plausible lowest-mass black hole ($\sim3.3\Msun$)~\citep{2019Sci...366..637T} and the possibly most massive NS so far in PSR J2215+5135 ($\sim2.27\Msun$)~\citep{2018ApJ...859...54L}.

The maximum mass of NSs, $M_{\rm TOV}$, has generated much interest due to its crucial importance on the formation of such systems~\citep[e.g.,][]{2021ApJ...908..106L,2020ApJ...899L..15S,2020ApJ...901L..34Y,2020ApJ...899L...1Z} and their merger rates~\citep[e.g.,][]{2020ApJ...899L...8F}, 
the detectability of produced kilonova~\citep[e.g.,][]{2020arXiv200906655D}, the mass distribution for NSs~\citep[e.g.,][]{2020PhRvD.101j3036G}, or discriminating between NSs and BHs through 
late inspiral 
to post-merger gravitational wave (GW) signal and electromagnetic (EM) counterparts~\citep[e.g.,][]{2013ApJ...766L..14H,2015ApJ...807L..24L,2015MNRAS.450L..85M,2018ApJ...856..110Y}, among many aspects.
Currently, it is still challenging to discriminate between these two populations~\citep[e.g.,][]{2020PhRvL.124g1101T}. 
Despite many recent works specifically discussing on the nature of GW190814~\citep[e.g.,][]{2021PhRvC.103b5808D,2021ApJ...908..122G,2021ApJ...908L...1T,2020ApJ...902...38Z,2020arXiv201111934Z,Miao2021} and the merger remnant of GW170817~\citep[e.g.,][]{2017ApJ...850L..19M,2018ApJ...852L..25R,2018PhRvD..97b1501R,2019PhRvD.100b3015S,2020ApJ...893..146A,2020PhRvL.125n1103B},
the conclusions have been diverse because of the unknown maximum mass and the uncertain equation of state (EOS) of NSs. While the Shapiro delay measurement of MSP J0740+6620~\citep{2020NatAs...4...72C} with 
mass $M=2.14^{+0.10}_{-0.09}\Msun$ (68\% confidence level) tightly constrained the lower bound on 
$M_{\rm TOV}$, 
its actual value is still a subject of debate. 
Theoretically, the EOS gives rise to a unique sequence of stellar configurations 
under hydrostatic equilibrium through the Tolman-Oppenheimer-Volkoff (TOV) equations, 
which naturally predicts a maximum mass for NSs.
As NS cores possess densities possibly up to $\approx 8-10\,n_0$ (where $n_0 = 0.16~\rm fm^{-3}$ is the nuclear saturation density), a description of the high-density core matter in terms of only hadrons and leptons might be inadequate~\citep{2020NatPh..16..907A}.
In addition, Bayesian analyses of NS observational data can lead to different results depending on whether phase transition is taken into account or not in the prior assumption of the EOS~\citep{2021PhRvL.126f1101A,2020arXiv200706526L}.
It is worth mentioning that the standard interpretation for the LIGO/Virgo and \nicer ~data 
are usually based on the parameterization of the EOS using a piecewise polytrope model, which does not explicitly include phase transitions. 

In this work we use Bayesian inference to infer $M_{\rm TOV}$ 
in the context of a strong phase transition from hadronic matter into quark matter 
inside dense cores of NSs,
by exploiting the binary tidal deformability ($\tilde\Lambda$) 
constraint from GW170817 by LIGO/Virgo~\citep{2017PhRvL.119p1101A} and
the mass ($M$) and radius ($R$) 
measurements of PSR J0030+0451 by \nicer~\citep{2019ApJ...887L..24M,2019ApJ...887L..21R}.
For the hadronic phase, we utilize two different EOSs with varying stiffness at higher densities~\citep{2016PhRvC..94c5804F,2018ApJ...862...98Z} while maintaining consistency with low-density laboratory nuclear experiments.
The main uncertainty arises from the quark phase EOS for which we treat using the generic parametrization based on sound velocity~\citep{2013PhRvD..88h3013A}. 
The phase transition parameters are then constrained to be compatible with the LIGO/Virgo and \nicer ~data.

This paper is organized as follows.
In Section 2, we introduce the unified EOSs for the hadronic phase of NSs
and the CSS parameterization 
for the hadron-quark transition, and the quark matter EOS. 
Section 3 presents the observational constraints 
employed and the Bayesian analysis 
method that we apply.
Our results are presented in Section 4 and summarized in Section 5.

\section{Neutron star equation of state}
\label{sec:eos}

We describe a static hybrid NS for which gravity is balanced by the pressure 
of the compressed dense matter encountered in its interior. 
From the NS surface to its core, with increasing density 
the following regions appear sequentially:\\
{\bf Outer~crust:} non-uniform Coulomb lattice of neutron-rich nuclei embedded in a degenerate electron gas (below the neutron drip density). 
The properties of NS outer crust are well constrained and we use the standard EOS proposed by~\citet{1971ApJ...170..299B};\\
{\bf Inner~crust:} non-uniform system composed of more 
exotic neutron-rich nuclei, degenerate electrons, and superfluid neutrons/superconducting protons (below the nuclear saturation density); \\
{\bf Outer~core:} uniform dense nuclear matter in weak-interaction equilibrium (or $\beta-$equilibrium) with leptons. 
We describe both 
the inner crust and the outer core 
using the unified EOSs with the same underlying nuclear interaction as explained in Sec.~\ref{sec:qmfdd2};\\
{\bf Inner~core:} uniform quark matter 
EOS represented by the CSS parameterization as explained in Sec.~\ref{sec:css}.\\

\subsection{Unified EOSs for nonuniform and uniform hadronic matter: QMF and DD2}
\label{sec:qmfdd2}

\begin{figure}
\centering
\includegraphics[width=3.2in]{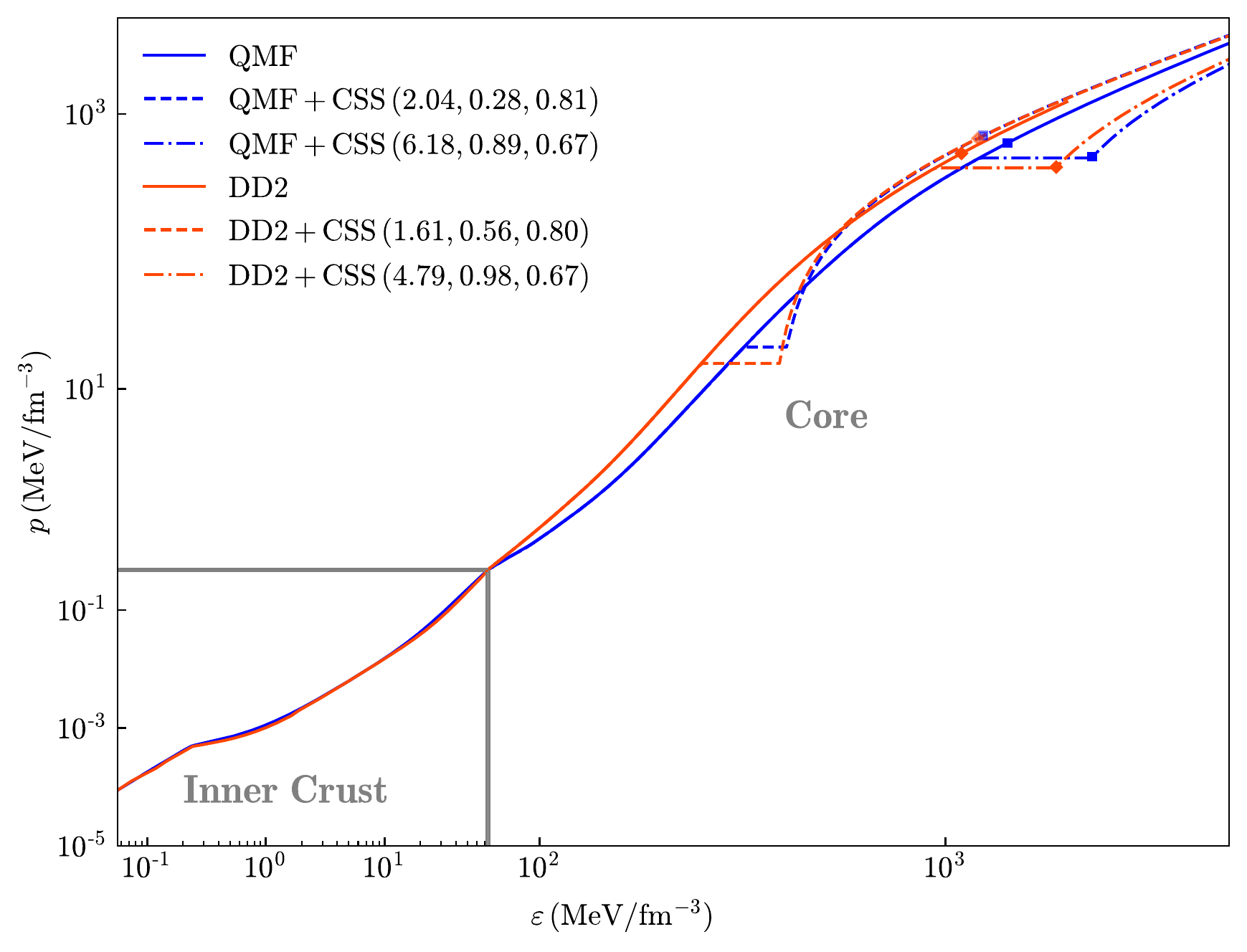}
\includegraphics[width=3.2in]{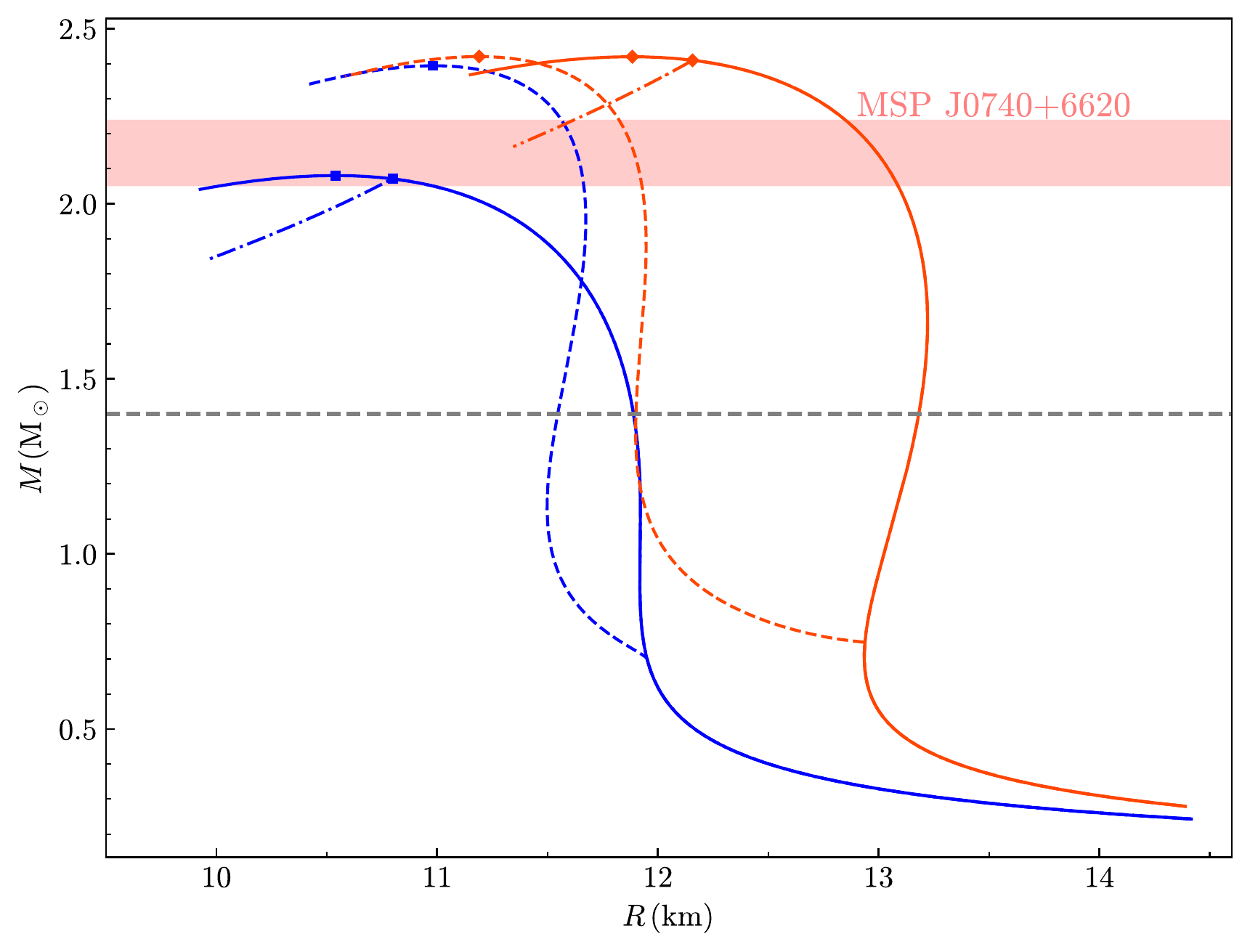}
\caption{
Upper panel: 
unified QMF (soft) and DD2 (stiff) hadronic EOSs employed in this study for the inner crust and outer core of NSs, along with four exemplary CSS parameter sets ($n_{\rm trans}/n_0, \Delta\varepsilon/\varepsilon_{\rm trans}, c_{\rm QM}^2$) for the inner quark-matter core. 
Lower panel: 
corresponding mass-radius relations for NSs with or without a quark core; 
the horizontal line indicates $M=1.4\Msun$.
Colored symbols (squares and diamonds) represent 
the central densities of the maximum-mass stars in each case. 
Also shown is the mass measurement of MSP J0740+6620 with $M=2.14^{+0.10}_{-0.09}\Msun$ (68\% confidence level)~\citep{2020NatAs...4...72C}. 
If the phase transition only occurs at sufficiently high densities, with the CSS parameters being $(6.18,0.89,0.67)$ for QMF and $(4.79,0.98,0.67)$ for DD2 as examples, 
the central densities of the maximum-mass hybrid stars are close to the initial density of the quark phase, indicating that only a small fraction of quark matter is present in the core 
(see more discussion in Appendix~\ref{sec:A1}).
}
\label{fig:eos}
\end{figure}

The QMF \citep{2018ApJ...862...98Z} and DD2~\citep{2016PhRvC..94c5804F} 
models are employed 
to represent 
fiducial soft and stiff 
hadronic matter, 
respectively.
To avoid uncertainties due to the crust, we utilize unified EOSs with a consistent treatment of NS crusts and the crust-core transition properties 
along with their cores. 
For NSs without a quark core, 
the \textit{soft} QMF EOS 
leads to a maximum mass 
$M_{\rm TOV}=2.07\Msun$ and a typical radius 
$R_{\rm 1.4}=11.77\,\rm km$,
whereas for 
the \textit{stiff} DD2 EOS 
$M_{\rm TOV}=2.42\Msun$ and $R_{\rm 1.4}=13.17\,\rm km$.
Both EOSs are constructed within the widely used relativistic mean-field (RMF) approach, based on an effective Lagrangian with meson fields mediating strong interactions between quarks or hadrons. 
Compatibility with available experimental 
as well as observational constraints at 
sub-nuclear and higher densities 
is ensured~\citep[e.g.,][]{2020PhRvL.125t2702D}.

\subsection{CSS parameterization for the hadron-quark phase transition ($n_{\rm trans}/n_0, \Delta\varepsilon/\varepsilon_{\rm trans}, c_{\rm QM}^2$)} 
\label{sec:css}

For high-density quark matter 
the ``constant-speed-of-sound'' (CSS) parameterization~\citep{2013PhRvD..88h3013A} is applied, and
the EOS from the onset of the phase transition up to the maximum central density of a star is
determined by three dimensionless parameters: the transition density $n_{\rm trans}/n_0$, the transition strength $\Delta\varepsilon/\varepsilon_{\rm trans}$, and the sound speed squared in quark matter $c^2_{\rm QM}$ (we work in units where $\hbar=c=1$).
The full EOS is 
\begin{eqnarray}
\varepsilon(p) = \left\{\!
\begin{array}{ll}
\varepsilon_{\rm HM}(p), & p<p_{\rm trans} \\ \nonumber 
\varepsilon_{\rm HM}(p_{\rm trans})+\Delta\varepsilon+c_{\rm QM}^{-2} (p-p_{\rm trans}), & p>p_{\rm trans}  
\end{array}
\right.
\end{eqnarray}
where $n_{\rm trans} \equiv n_{\rm HM}(p_{\rm trans})$ and $\varepsilon_{\rm trans} \equiv \varepsilon_{\rm HM}(p_{\rm trans})$, and $\Delta\varepsilon/\varepsilon_{\rm trans}$ is essentially the finite discontinuity in the energy density at the phase boundary.
In practice, our study is limited to first-order phase transitions with a sharp interface (Maxwell construction).

Since there exists a one-to-one correspondence between 
the underlying EOS 
and global properties of NSs such as ($M,R,\Lambda$), 
predicted values 
for these quantities 
can therefore be 
characterized by 
the three CSS 
parameters ($n_{\rm trans}/n_0, \Delta\varepsilon/\varepsilon_{\rm trans}, c_{\rm QM}^2$) 
in the hybrid EOSs. 
In Fig.~\ref{fig:eos} we show four 
parameter sets as examples 
for the high-density quark matter 
together with the unified QMF and DD2 EOSs for the inner crust and the outer core. 
Before reporting our results, 
we first 
introduce the scheme of the Bayesian inference approach that allows direct and consistent utilization of the LIGO/Virgo and \nicer ~observational data.

\section{Observational constraints and Bayesian analysis}
\label{sec:analysis}

According to Bayes theorem, the posterior probability density function (PDF) of a set of parameters $\vec\theta$ given a data $d$ is expressed as
\begin{equation}
    p(\vec\theta|d) = \frac{p(d|\vec\theta)p(\vec\theta)}{\int p(d|\vec\theta)p(\vec\theta)d\vec\theta}.
\end{equation}
Here $p(d|\vec\theta)$, usually written in $\mathcal{L}(d|\vec\theta)$, is the likelihood function of a data $d$ given a set of parameters $\vec\theta$, and $p(\vec\theta)$ is the prior PDF of the parameters, which reflects our preliminary knowledge of the parameters. The denominator is the evidence for given data $d$ and can be treated as a normalization factor. We employ the Python {\sc Bilby} \citep{2019ascl.soft01011A} package to sample the parameters and calculate the marginal likelihoods. 
For the sampler we choose the {\sc Pymultinest} \citep{2016ascl.soft06005B} code which is based on a nested sampling algorithm. The credible intervals (regions) of one (two) parameter(s) can then be obtained by marginalizing over all other parameters.

\begin{figure*}
\centering
\includegraphics[width=3.4in]{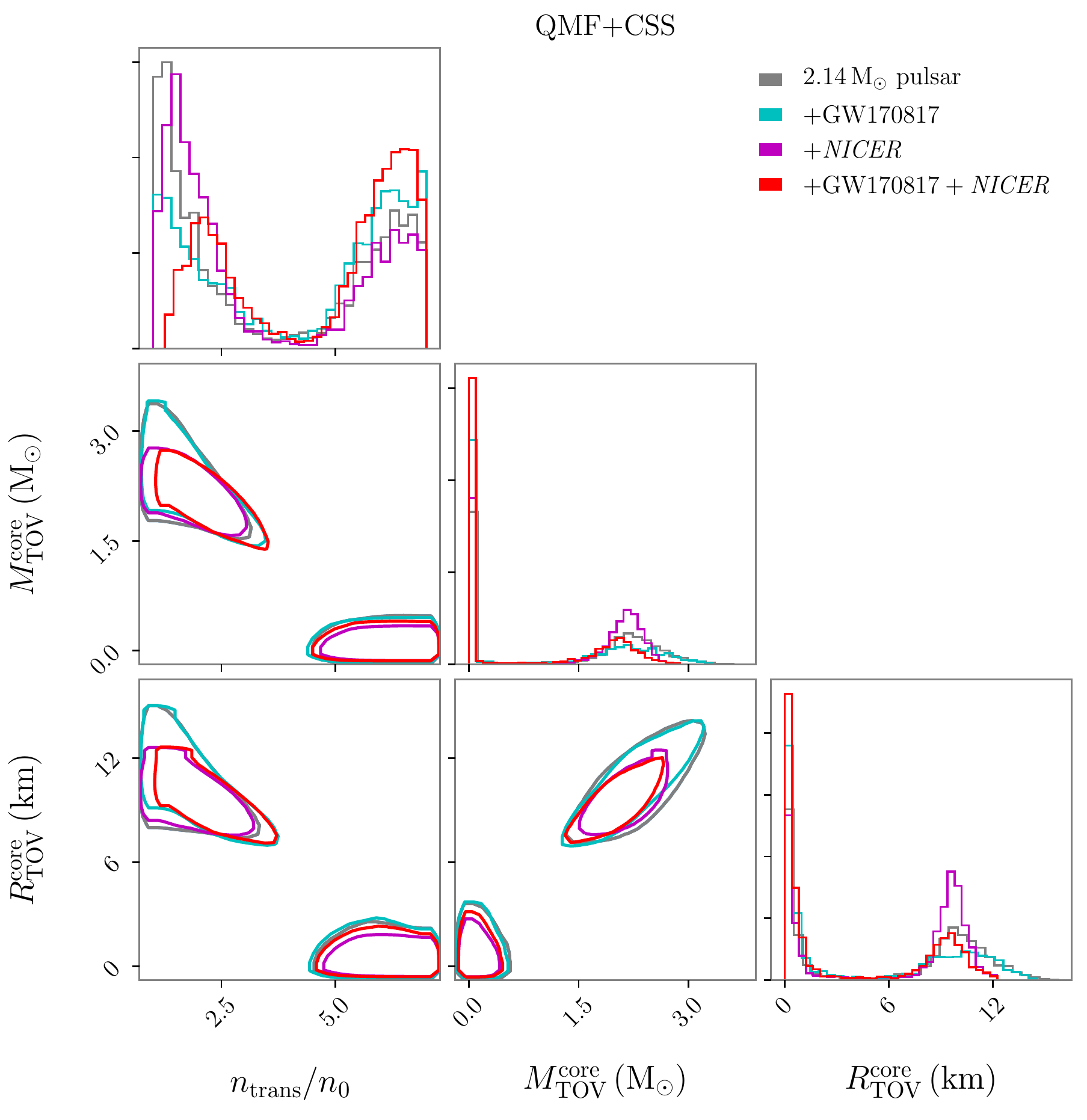}
\hspace{+5mm}
\includegraphics[width=3.4in]{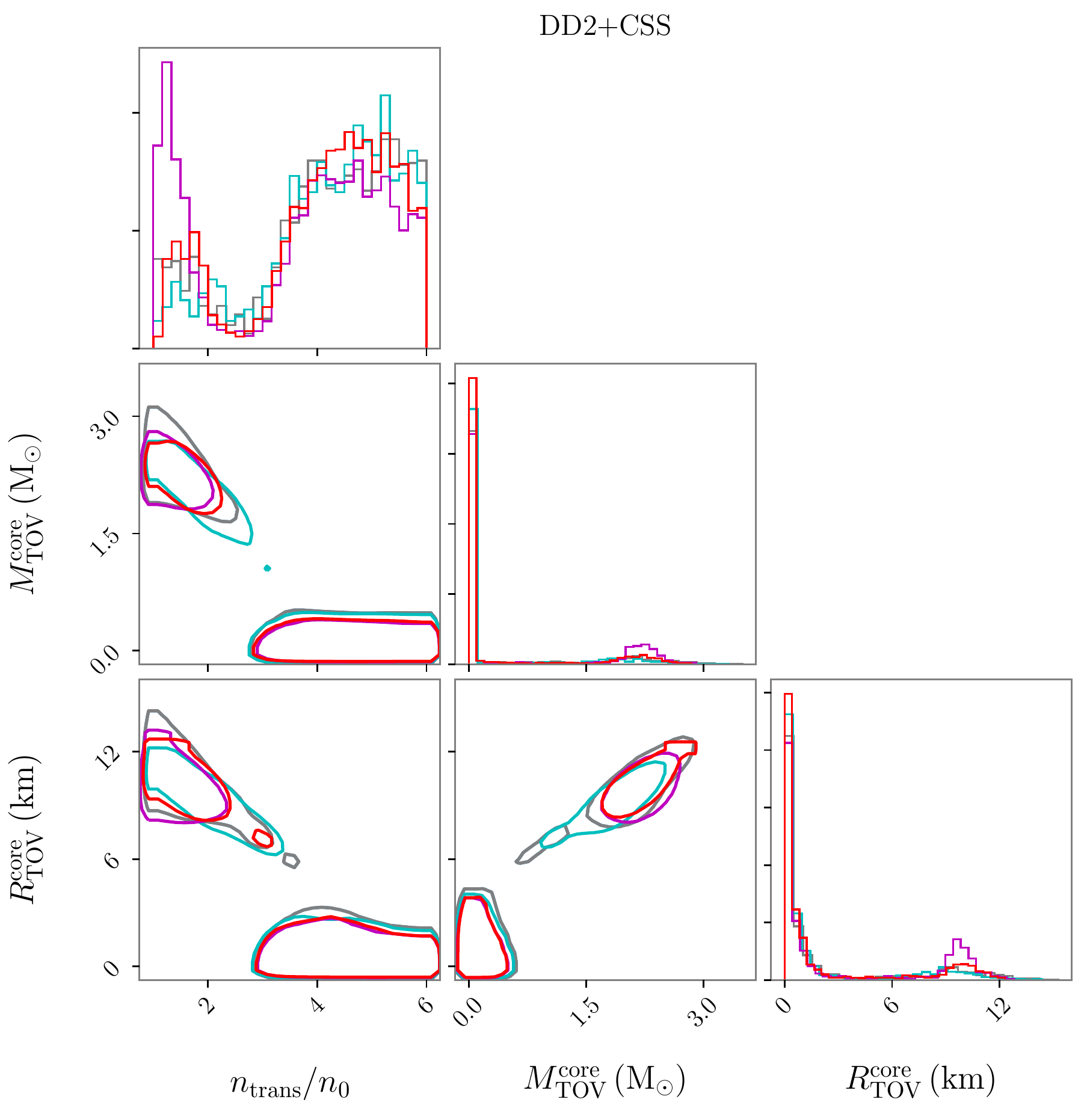}
\caption{Posterior distributions of the transition density (in units of the saturation density) $n_{\rm trans}/n_0$ and the mass and radius of quark matter core ($M^{\rm core}_{\rm TOV}$, $R^{\rm core}_{\rm TOV}$) for hybrid stars within the frameworks of QMF+CSS (left) and DD2+CSS (right). 
The contours are the $90\%$ credible regions for these parameters.
The results are conditioned on the prior of four different analyses (see details in Sec.~\ref{sec:priors_constraints}).
For both soft QMF and stiff DD2 hadronic EOSs, there are two scenarios for the maximum-mass hybrid stars: with a small ($R_{\rm core} \ll \rm 6~km$) or a large ($R_{\rm core} \gg \rm 6~km$) quark core.
Within QMF+CSS, the large (small) core case corresponds to $n_{\rm trans}/n_0\ll4$ ($n_{\rm trans}/n_0\gg4$).
Within DD2+CSS, the large (small) core case corresponds to $n_{\rm trans}/n_0\ll3$ ($n_{\rm trans}/n_0\gg3$).
} 
\label{fig:multimodal_posterior}
\end{figure*}

\subsection{Likelihoods and constraint}
\label{sec:likelihoods}
In the present analysis, the likelihood is defined as 
\begin{equation}
    \mathcal{L}(d|\vec\theta)=\mathcal{L}_{M_s}\times\mathcal{L}_{\rm GW}\times\mathcal{L}_{\rm PSR}
\end{equation}
where $\mathcal{L}_{M_s}$ is the likelihood to encapsulate the mass measurement of a pulsar source. $\mathcal{L}_{\rm GW}$ and $\mathcal{L}_{\rm PSR}$ are the likelihoods of the GW170817 data and the mass-radius measurement of PSR J0030+0451 by \nicer, respectively. If we only focus on the GW (PSR) data, we will neglect the PSR (GW) data.

{\bf Lower bound on $M_{\rm TOV}$ from \objectname{MSP J0740+6620}.} The mass measurements of massive pulsars establish a firm lower bound on the NS maximum mass. Only the EOSs that 
support a $M_{\rm TOV}$ larger than this lower bound can pass this constraint, while others will be rejected.
Assuming the mass measurement is a Gaussian distribution N$(\mu,\sigma^2)$, the likelihood can be written as 
\begin{equation}
    \mathcal{L}_{M_s} = \Phi(\frac{M_{\rm TOV}(\theta)-\mu}{\sigma})
\end{equation}
where $\Phi(x)\equiv\int_{-\infty}^{x}(2\pi)^{-\frac12}e^{-\frac{x^2}{2}}dx$ is the standard Gaussian cumulative distribution function.
In the present analysis, we choose the highest mass measured through Shapiro delay, $M=2.14_{-0.09}^{+0.10}\Msun$ (68\% confidence level) of \objectname{MSP J0740+6620}~\citep{2020NatAs...4...72C}, to place the $M_{\rm TOV}$ constraint. 
We fit the mass distribution of this source by a Gaussian distribution with $\mu = 2.14\,\Msun$ and $\sigma = 0.1\,\Msun$. 

{\bf GW170817.} The first binary NS merger GW170817 detected by the LIGO/Virgo collaboration~\citep{2017PhRvL.119p1101A} provided significant insight into the 
matter EOS and the NS internal structure because of 
their effects on the gravitational waveform. Assuming that the noise in the LIGO/Virgo detectors is Gaussian and stationary, the functional form of likelihood is often expressed as 
\begin{equation}
\label{eq:Likelihood}
\mathcal{L}_{\rm GW}\!\propto\!{\rm exp}\!\left[-2 \int_{0}^{\infty} \frac{|\tilde d(f) - \tilde h(f,\vec{\theta}_{\rm GW})|^2}{S_{\rm n}(f)} df\right],
\end{equation}
where $S_{\rm n}(f)$, $\tilde d(f)$, and $\tilde h(f,\vec{\theta}_{\rm GW})$ denote the power spectral density (PSD), the Fourier transform of the measured strain data, and the frequency domain waveform generated using the parameter set $\vec{\theta}_{\rm GW}$, respectively. 
In the present analysis we adopt a 128s strain data\footnote{\url{https://www.gw-openscience.org/eventapi}} from GPS time 1187008682s to 1187008890s and the PSDs\footnote{\url{https://doi.org/10.7935/KSX7-QQ51}} of GW170817 \citep{2019PhRvX...9c1040A} released by the LIGO/Virgo collaboration. We also take the waveform model {\sc IMRPhenomD\_NRTidal} \citep{2017PhRvD..96l1501D} for illustrative purposes. 

\begin{table*}
  \centering 
  \caption{Most probable intervals of various phase transition parameters (in the $2^{\rm nd}-4^{\rm th}$ 
  columns) and properties for the maximum-mass hybrid stars (in the $5^{\rm th}-7^{\rm th}$ columns), to the $90\%$ confidence level, constrained by four different analyses explained in Sec.~\ref{sec:priors_constraints}.
  $n_{\rm trans}/n_0$ is the hadron-quark phase transition density. $\Delta\varepsilon/\varepsilon_{\rm trans}$ is the discontinuity in the energy density where $n_{\rm trans} \equiv n_{\rm HM}(p_{\rm trans})$ and $\varepsilon_{\rm trans} \equiv \varepsilon_{\rm HM}(p_{\rm trans})$.
  $c_{\rm QM}^2$ is the squared speed of sound in the high-density quark phase. 
   $M_{\rm TOV}$ is the maximum mass, and $R_{\rm TOV}$ and $n^c_{\rm TOV}$ are the corresponding radius and central density, respectively.
   The results are shown for both the soft QMF hadronic EOS and the stiff DD2 one.
  }
  \setlength{\tabcolsep}{0.8pt}
\renewcommand\arraystretch{1.2}
 \vskip-3mm
  \begin{ruledtabular}
  \begin{tabular}{lccccc}      
    Parameters & &  $2.14 \Msun$ pulsar & +GW170817 & +\nicer & +GW170817+\nicer  \\
    \hline \hline 
       \multirow{2}{*}{$n_{\rm trans}/n_0$} & QMF &  $1.50_{-0.45}^{+1.69}$  & $1.57_{-0.49}^{+1.13}$ 
        &  $1.69_{-0.64}^{+1.59}$ & $2.04_{-0.83}^{+1.22}$ \\
     & DD2 &  $1.54_{-0.49}^{+1.29}$  & $1.40_{-0.35}^{+0.84}$ 
        &  $1.92_{-0.79}^{+0.97}$ & $1.61_{-0.46}^{+0.92}$ \\
    \hline
    \multirow{2}{*}{$\Delta\varepsilon/\varepsilon_{\rm trans}$} & QMF  &  $0.39_{-0.35}^{+1.03}$  & $0.65_{-0.55}^{+0.79}$ &  $0.23_{-0.21}^{+0.58}$  &  $0.28_{-0.25}^{+0.39}$ \\
    & DD2 &  $0.38_{-0.34}^{+1.04}$  & $0.93_{-0.69}^{+0.62}$ &  $0.32_{-0.28}^{+0.63}$  &  $0.56_{-0.43}^{+0.44}$ \\
     \hline
    \multirow{2}{*}{$c_{\rm QM}^2$} & QMF &  $0.80_{-0.32}^{+0.17}$   &  $0.82_{-0.24}^{+0.16}$  & $0.77_{-0.33}^{+0.21}$ &  $0.81_{-0.28}^{+0.17}$ \\
    & DD2 &  $0.78_{-0.34}^{+0.19}$ &  $0.82_{-0.25}^{+0.16}$ & $0.80_{-0.35}^{+0.18}$ &  $0.80_{-0.29}^{+0.18}$ \\
    \hline \hline
   \multirow{2}{*}{$M_{\rm TOV}/ \Msun$} & QMF &  $2.38_{-0.30}^{+0.74}$   &  $2.31_{-0.22}^{+0.33}$  & $2.42_{-0.34}^{+0.68}$ &  $2.36_{-0.26}^{+0.49}$ \\
   & DD2 &  $2.42_{-0.33}^{+0.72}$   &  $2.30_{-0.22}^{+0.31}$  & $2.39_{-0.31}^{+0.60}$ &  $2.39_{-0.28}^{+0.42}$ \\
    \hline
     \multirow{2}{*}{$R_{\rm TOV}/\rm km$}  & QMF  &  $11.09_{-1.42}^{+2.99}$   &  $10.57_{-0.84}^{+1.52}$ & $11.48_{-1.47}^{+2.34}$ &  $10.96_{-1.01}^{+1.79}$ \\
     & DD2  &  $11.66_{-1.73}^{+2.62}$ &  $10.62_{-0.81}^{+1.48}$ & $11.53_{-1.40}^{+1.90}$ &  $11.24_{-1.10}^{+1.42}$ \\
    \hline
    \multirow{2}{*}{$n^c_{\rm TOV}/\rm fm^{-3}$} & QMF  &  $0.93_{-0.36}^{+0.30}$  &  $1.02_{-0.24}^{+0.21}$  & $0.88_{-0.30}^{+0.30}$ &  $0.96_{-0.27}^{+0.24}$ \\
    & DD2   &  $0.87_{-0.33}^{+0.32}$  &  $1.02_{-0.23}^{+0.19}$  & $0.90_{-0.28}^{+0.27}$ &  $0.93_{-0.23}^{+0.22}$ \\
  \end{tabular}
  \end{ruledtabular}
  \label{tb:1}
\end{table*}

{\bf \objectname{PSR J0030+0451}.} The recent simultaneous mass-radius 
measurements 
of PSR J0030+0451 from \nicer, $M=1.44^{+0.15}_{-0.14} \Msun$, $R=13.02^{+1.24}_{-1.06}~\rm km$~\citep{2019ApJ...887L..24M}
and $M=1.34^{+0.15}_{-0.16} \Msun$, $R=12.71^{+1.14}_{-1.19}~\rm km$~\citep{2019ApJ...887L..21R}, to the $68.3\%$ credibility interval, provide an additional useful constraint on the NS EOS. 
Since results in 
\citet{2019ApJ...887L..21R} and \citet{2019ApJ...887L..24M} 
are consistent with each other, we select the best fit scenario within the ST+PST model of \citet{2019ApJ...887L..21R}. 
In this case, the likelihood of generating the mass-radius distribution of \objectname{PSR J0030+0451} is expressed as 
\begin{equation}
    \mathcal{L}_{\rm PSR} = {\rm KDE}(M, R\mid \vec{S}),
\end{equation}
where the right-hand side is a Gaussian Kernel Density Estimation (KDE) of the posterior samples $\vec{S}$ of the mass and radius given by \citet{2019ApJ...887L..21R}.

\subsection{Parameters and priors}
\label{sec:priors_constraints}

The parameters related to the hybrid star EOSs are $\vec\theta_{\rm EOS} = \{n_{\rm trans}/n_0,\Delta\varepsilon/\varepsilon_{\rm trans},c_{\rm QM}^2\}$ 
within the CSS parameterization. 
Given that a normal 
NS reaches the maximum-mass configuration
with its central density
below $\approx7.0\,n_0$/$6.0\,n_0$ 
for the QMF/DD2 hadronic EOS, 
we choose a uniform distribution of $n_{\rm trans}/n_0$ for hybrid stars 
in the range of $[1,7]$/$[1,6]$, respectively. 
For the sound speed squared in quark matter $c_{\rm QM}^2$, we vary its value from $c_{\rm QM}^2=1/3$ (the conformal limit in perturbative QCD matter) to $c_{\rm QM}^2=1$ (the causal limit), with a uniform distribution. 
We also assign a reasonably wide 
range for the energy density discontinuities as $\Delta\varepsilon/\varepsilon_{\rm trans} \in[0, 2]$ with a uniform distribution. 

For GW170817, the GW parameters are chirp mass in the detector frame, mass ratio, tidal deformabilities, aligned spins, inclination angle, geocentric time, polarization of GW, orbital phase at coalescence, red-shift of the source, and its position in the sky, i.e., $\vec\theta_{\rm GW} = \{\mathcal{M}^{\rm det}, q, \Lambda_1,\Lambda_2,\chi_{1z},\chi_{2z},\theta_{\rm jn},t_{\rm c},\Psi,\varphi, z,\alpha,\delta\}$. 
We fix the location of the source to the position determined by EM observations \citep{2017ApJ...848L..12A,2017ApJ...848L..28L} with $\alpha\,({\rm J}2000.0)=13^{\rm h}09^{\rm m} 48^{\rm s}.085$, $\delta\,({\rm J}2000.0)=-23^\circ22^\prime53^{\prime\prime}.343$ and $z=0.0099$. 
$\mathcal{M}^{\rm det}$, $q$, $\chi_{1z}(\chi_{2z})$, $t_c$ and $\Psi$ are chosen to distribute uniformly in the range $[1.18,1.21]M_\odot$, $[0.5,1.0]$, $[-0.05,0.05]$, $[1187008882,1187008883]$s, and $[0,2\pi]$. 
For the inclination angle $\theta_{\rm jn}$,  we assume a uniform prior distribution.\footnote{Both the GW data \citep{2017PhRvL.119p1101A} and the EM data \citep{2018Natur.561..355M} as well as structured jet modeling \citep{2020MNRAS.498.5643T} can give some constraints on the viewing angle for GW170817. However, since the constrained viewing angle in these studies gives a very wide distribution, the uniform prior distribution can serve for the purpose of our study.}
The tidal deformabilities 
$\Lambda_1$ and $\Lambda_2$ 
are related to the component-star masses through the hybrid star EOS, i.e.,
\begin{equation}
\begin{split}
    &\Lambda_1 = \Lambda(\vec\theta_{\rm EOS}, m_1), \\
    &\Lambda_2 = \Lambda(\vec\theta_{\rm EOS}, m_2),
\end{split}
\end{equation}
where the 
component-star 
masses $m_1$, $m_2$ can be obtained with
\begin{equation}
\begin{split}
     &m_1 = \mathcal{M}^{\rm det}(1+q)^{1/5}/q^{3/5}/(1+z), \\
     &m_2 = m_1q.
\end{split}
\end{equation}
To improve the convergence rate of the nest sampler, we marginalize the likelihood over the orbital phase at coalescence. 

The mass and radius entering the likelihood of PSR data are related to the hybrid star EOS via the central pressure of the star, $p_c$, i.e.,
\begin{equation}
    \begin{split}
        &M=M(\vec\theta_{\rm EOS}, p_c),\\
        &R=R(\vec\theta_{\rm EOS}, p_c).
    \end{split}
\end{equation}
Therefore we only need one additional parameter $p_c$ to perform the analysis if we take account of the PSR data.

\begin{figure*}
\centering
\includegraphics[width=3.2in]{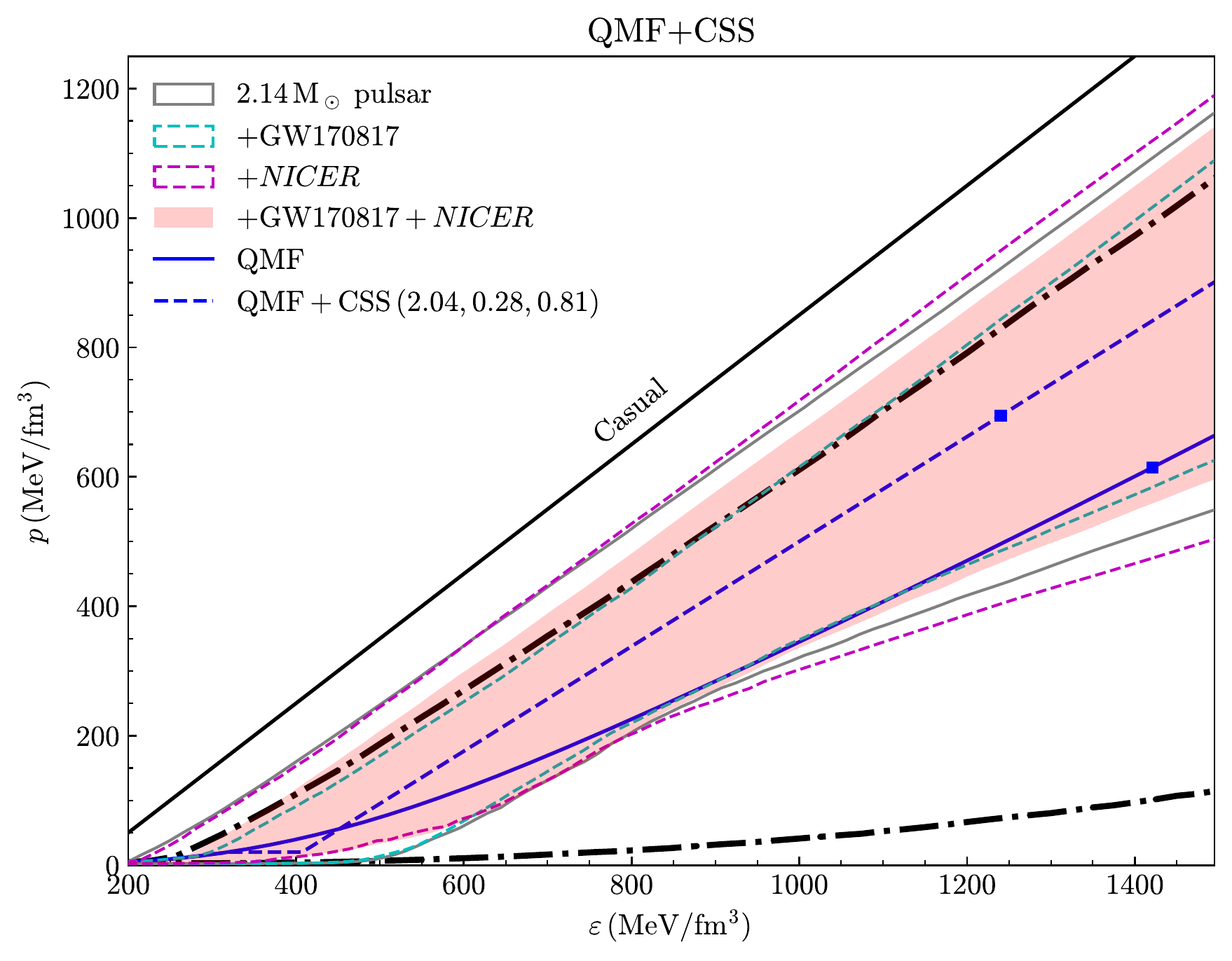}
\hspace{+5mm}
\includegraphics[width=3.2in]{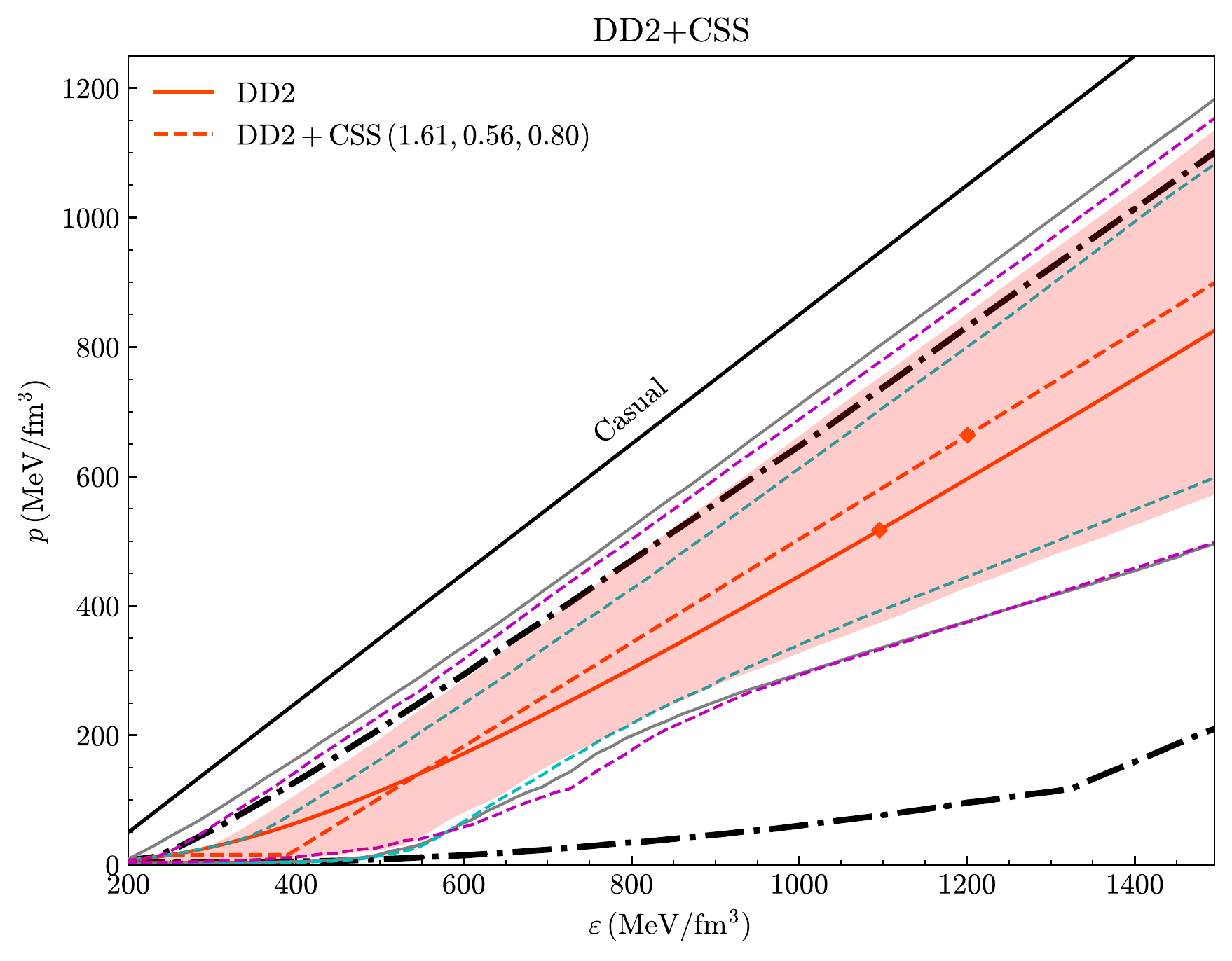}
\caption{Posterior distributions of the hybrid EOS: 
the grey, blue, purple, and pink lines (or shaded regions) show the $95\%$ credible regions conditioned on the prior (indicated with bold dash-dotted lines) of four different analyses (see details in Sec.~\ref{sec:priors_constraints}); 
Shown together are the results from the pure hadronic EOS as well as two exemplary hybrid stars using QMF/DD2+CSS ($n_{\rm trans}/n_0, \Delta\varepsilon/\varepsilon_{\rm trans}, c_{\rm QM}^2$), where colorful symbols indicate central densities of the maximum-mass stars, respectively. 
The black lines represent the extreme causal EOSs. 
}
\label{fig:multi_posterior_eos}
\end{figure*}

We carry out four main tests to investigate how each data set 
(see details in Sec.~\ref{sec:likelihoods}) would affect the result, namely: \\
(i) $2.14 \Msun$ pulsar: where we apply the minimum $M_{\rm TOV}$ constraint from \objectname{MSP J0740+6620}; \\
(ii) +GW170817: where we consider 
both 
the constraint in (i) and the GW data of GW170817; \\ 
(iii) +\nicer: where we consider 
both the constraint in (i) and the \nicer ~measurement of PSR J0030+0451; \\ 
(iv) +GW170817+\nicer: where we combine the data of PSR J0030+0451 and GW170817 all together with the maximum-mass constraint in (i).

\section{Results and discussion}
\label{sec:res}

Our analysis first reveals that there are two scenarios for the maximum-mass hybrid stars: 
with a small ($R_{\rm core} \ll \rm 6~km$) or large ($R_{\rm core} \gg \rm 6~km$) quark core, corresponding to an early or late onset of the quark phase in the hybrid EOS, respectively.
As detailed in Fig.~\ref{fig:multimodal_posterior}, 
this conclusion is not subject to the hadronic EOS used or the choice of including or excluding any individual data. 
Since the small-core hybrid stars have a high threshold density for quark matter to appear which is close to the central density of the maximum-mass hybrid stars 
(see e.g. the dot-dashed curves in Fig.~\ref{fig:eos}), only a small fraction of the NS is made of 
quark matter. 
This results in a limited or even negligible effect on the EOS and a similar maximum mass for hybrid stars compared to that of pure hadronic stars.

We further compute the Bayes Factor of the hybrid EOS with respect to the purely hadronic one to quantify this effect, and indeed find its value in general close to 1, e.g. $B=1.486~(1.782)$ for QMF (DD2) within the joint GW170817+{\it NICER} analysis.
Consequently, in the following analysis we only focus on the large-core hybrid stars because it allows a much larger parameter space for the 
NS masses and radii compared to the small-core case~\citep{2020ApJ...904..103M}.
To do this, hereafter we reset the prior of $n_{\rm trans}/n_0$ to be in the range of $[1,4]$/$[1,3]$ for the large-core case within QMF/DD2+CSS, while for the small-core case the range is $[4,7]$/$[3,6]$. 
In Appendix \ref{sec:A1}, we report supplementarily detailed results for the small-core hybrid stars in the QMF hadronic case.

\subsection{Hybrid star EOS and the maximum mass}

\begin{figure*}
\centering
\includegraphics[width=7.0in]{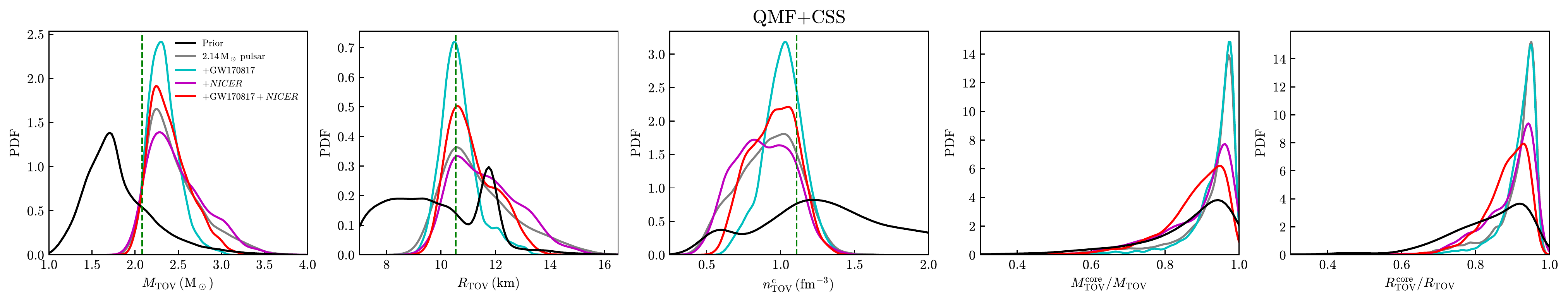}
\includegraphics[width=7.0in]{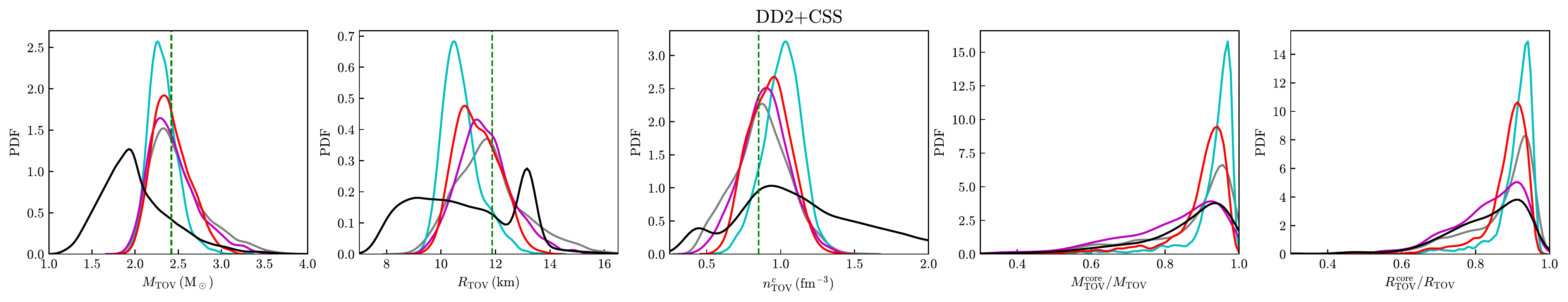}
\caption{Posterior PDFs of various properties ($M_{\rm TOV}, R_{\rm TOV}, n_{\rm TOV}^c, M^{\rm core}_{\rm TOV}/M_{\rm TOV}, R^{\rm core}_{\rm TOV}/R_{\rm TOV}$) for the maximum-mass hybrid stars (90\% confidence level) conditioned on the priors (shown in black lines in each panel) of four different analyses (see details in Sec.~\ref{sec:priors_constraints}).
The green vertical lines in the left six panels denote the corresponding results of NSs without a quark core.
} 
\label{fig:multi-posterior of NS prop for big core case}
\end{figure*}

Figure~\ref{fig:multi_posterior_eos} illustrates the posterior distributions of the hybrid EOS, 
and the most probable values for the phase transition parameters 
and their $90\%$ confidence boundaries
are summarized in Table~\ref{tb:1}, $2^{\rm nd}$ to $4^{\rm th}$ columns.
The $5^{\rm th}-7^{\rm th}$ columns of Table~\ref{tb:1} 
list the corresponding results for three properties of 
maximum-mass hybrid stars ($M_{\rm TOV}, R_{\rm TOV}, n_{\rm TOV}^c$). 
In Fig.~\ref{fig:multi-posterior of NS prop for big core case}, we report in details the posterior PDFs of five stellar configuration parameters ($M_{\rm TOV}, R_{\rm TOV}, n_{\rm TOV}^c, M^{\rm core}_{\rm TOV}/M_{\rm TOV}, R^{\rm core}_{\rm TOV}/R_{\rm TOV}$).
See also Appendix~\ref{sec:A2} for similar predictions for 
$1.4\Msun$ and $2.0\Msun$ hybrid stars. 

It is commonly accepted that the inclusion of extra degrees of freedom (such as deconfined quarks discussed in the present work) tends to soften the NS matter EOS. 
However, as long as the high-density phase is sufficiently stiff i.e. with a large enough sound speed, 
through the introduction of a hadron-quark phase transition the parameter space for NS EOSs 
could be extended significantly~\citep{2020ApJ...904..103M} (see e.g. colored 
solid curves and pink shaded regions 
in Fig.~\ref{fig:multi_posterior_eos}), 
especially for soft hadronic EOSs such as QMF. 
In fact, the QMF hadronic 
EOS is quite soft due to the inclusion of the quark interaction for describing the inner structure of a nucleon in vacuum; see details in a recent review~\citep{2020JHEAp..28...19L}. 
As a result, $M_{\rm TOV}$ of hybrid NSs with a quark core is larger/smaller (thus the corresponding central density is relatively smaller/larger) 
than that of normal NSs based on the soft QMF/stiff DD2 hadronic EOS, 
falling in a relatively robust range of $\sim2.10-2.85\Msun$ (90\% confidence level). 
The probability distribution of quark matter for the maximum-mass star peaks within a sphere of $R^{\rm core}_{\rm TOV}/R_{\rm TOV}\approx0.90$ with its mass fraction larger than $90\%$, corresponding to the scenario of a \textit{big} quark core. 

From Table~\ref{tb:1} one can see that 
the joint analysis of the GW170817 and \nicer ~data 
prefers the hadron-quark phase transition taking place at not-too-high densities ($n_{\rm trans}/n_0\approx2.04$ for QMF and $1.61$ for DD2, respectively). Such a low threshold is consistent with previous studies~\citep{2021PhRvL.126f1101A,2021PhRvC.103c5802X}, 
while the sound speed squared $c_{\rm QM}$ above the transition should be larger than $0.53$ ($0.51$) for QMF (DD2), 
indicating a rather stiff quark EOS generically. 
The current statistical analysis is also consistent with previous model calculations which confirm that 
to ensure a large mass for hybrid stars above two solar masses, the core sound speed needs to be necessarily large~\citep{2019arXiv190600826X}, even when the assumption of a first-order phase transition
is relaxed~\citep{2020arXiv200906441D}.

\subsection{Astrophysical implications}

We now turn to examine the nature and properties of several merger events based on our results of the hybrid star maximum mass.
Even though there is a wide distribution of $M_{\rm TOV}$ from our results, the peak value $\sim2.4\Msun$ suggests that the possibility that the secondary component of GW190814 with a mass $M\sim2.6\Msun$~\citep{2020ApJ...896L..44A} could be a spinning supermassive NS is not ruled out, 
even if one takes into account the limitation of the intrinsic r-mode instability~\citep{2020arXiv201111934Z}. 
Within this scenario, however, the supramassive NS needs to carry a low dipolar magnetic field so that it has not been spun down yet at the time of merger. A more likely situation is that this secondary component is a black hole, since the typical spin-down timescale is quote short \citep{2020PhRvD.101f3021S}.
Considering the maximal spin observed in Galactic NSs, \citet{2019PhRvX...9a1001A} inferred that the 90\% credible intervals for the component masses of the GW170817 event lie between $1.16$ and $1.60\Msun$ (with the most probable total mass around $M_{\rm tot}=2.73\Msun$), and \citet{2020ApJ...892L...3A} reported the corresponding component masses ranging from $1.46$ to $1.87 \Msun$ (with the most probable total mass around $M_{\rm tot}=3.4\Msun$) for the GW190425 event. 
Our results on the maximum mass of NSs (up to $\sim2.85\Msun$ to the 90\% confidence level) suggest that the remnant of GW170817 could be 
a massive rotating NS \citep{2020ApJ...893..146A}. 
In comparison, the GW190425 event may be more compatible with a BH remnant,
especially given that it has no detection of a luminous counterpart,  
and the inferred total mass is $0.45\Msun$ above the most massive observed binary NS system (PSR J1913+1102).
Further studies on the stellar evolution are warranted to clarify these important issues. 
See Appendix~\ref{sec:A3} for our results using the GW190425 data.

The scenario that GW170817 left behind a massive NS has implications in interpreting the EM counterparts of binary NS mergers. First, since GW170817 was associated with the short GRB 170817A, it suggests that the launch of a GRB jet does not require the formation of a black hole to begin with. The argument that the existence of a short GRB necessitates the formation of a black hole, and hence, a small $M_{\rm TOV}$ \citep[e.g.][]{2017ApJ...850L..19M,2018ApJ...852L..25R,2018PhRvD..97b1501R}, is flawed. Rather, it suggests that there might exist an underlying long-lived engine that powers the late EM counterpart, including the kilonova \citep{2018ApJ...861L..12L,2018ApJ...861..114Y} and late X-ray emission \citep{2019MNRAS.483.1912P,2020MNRAS.498.5643T}. The 1.7 s delay between GRB 170817A and GW170817 was not due to the delay time to form a black hole, but rather due to the time delay for a promptly launched jet to reach the energy dissipation radius \citep{2018NatCo...9..447Z,2019FrPhy..1464402Z}. The lack of significant energy injection in the GW170817/GRB 170817A remnant required that the dipolar magnetic field strength is below $10^{13}$ G \citep{2018ApJ...860...57A,2020ApJ...893..146A}. Alternatively, most of the spin energy of the post-merger massive NS could have been carried away via secular gravitational waves \citep{2013PhRvD..88f7304F,2016PhRvD..93d4065G} or deconfined quarks \citep{2016PhRvD..93j3001D,2016PhRvD..94h3010L,2020PhRvD.101f3021S}. For a review on the post-merger remnant of binary NSs and GW170817 in particular, see \cite{2020GReGr..52..108B,2020arXiv201208172S}. The existence of a long-lived NS engine for short GRBs is consistent with the interpretation of the X-ray plateau commonly existing in short GRB afterglows \citep{2010MNRAS.409..531R,2013MNRAS.430.1061R,2015ApJ...805...89L} and the X-ray transients CDF-S XT2 and XT1 \citep{2019Natur.568..198X,2019ApJ...886..129S}, which pointed towards a $M_{\rm TOV}$ in the range of $\sim 2.4 \Msun$~\citep{2014PhRvD..89d7302L,2016PhRvD..93d4065G,2016PhRvD..94h3010L,2018MNRAS.481.3670R}. 
According to this picture, a large fraction of binary NS mergers will be accompanied by a bright X-ray emission component \citep{2013ApJ...763L..22Z}. Future regular monitoring of X-ray counterparts from binary NS merger GW sources with e.g. Einstein Probe \citep{2018SSPMA..48c9502Y}, will test the scenario of a relatively large $M_{\rm TOV}$ suggested here.

\begin{figure*}
\centering
\includegraphics[width=7in]{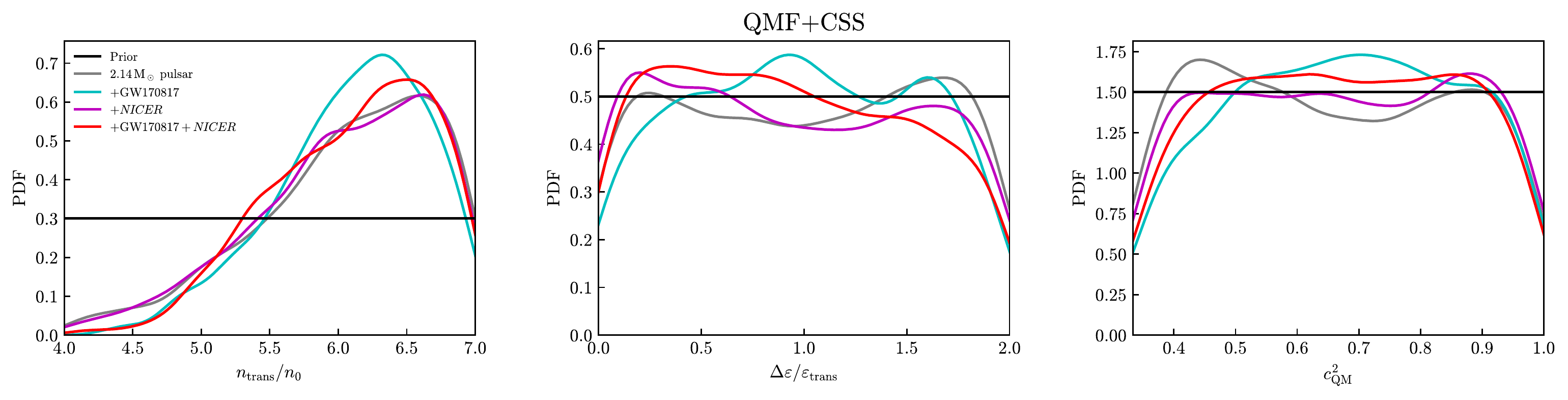}
\includegraphics[width=7in]{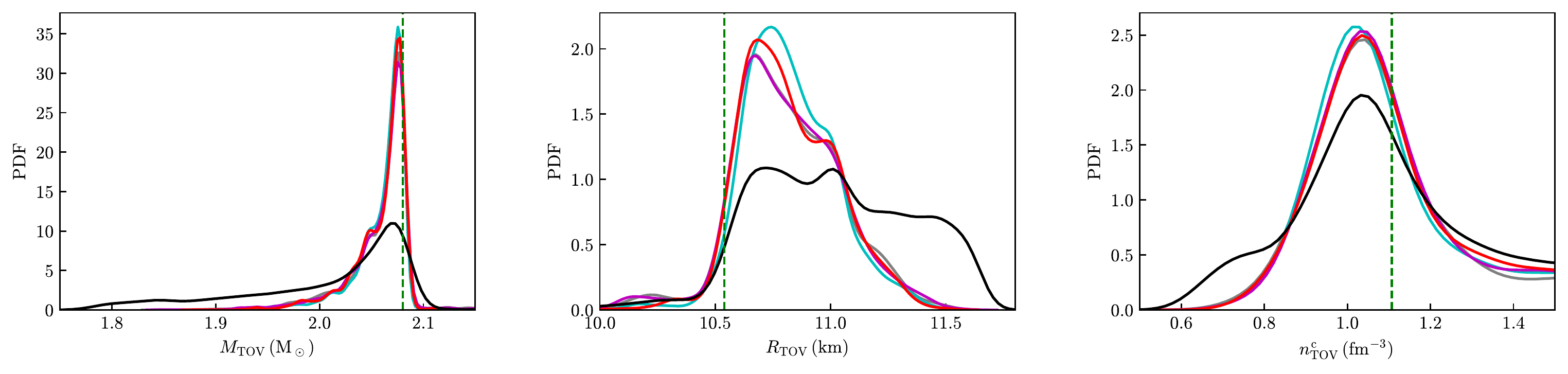}
\caption{Posterior PDFs of the CSS parameters (upper panels) and the maximum-mass hybrid star properties (lower panels), to the 90\% confidence level, for the small core case (i.e., $n_{\rm trans}\gtrsim 4\,n_0$) in Fig.~\ref{fig:multimodal_posterior}.
The analysis is performed within the QMF+CSS framework conditioned on the priors (shown in black lines in each panel) of four different analyses (see details in Sec.~\ref{sec:priors_constraints}). 
The green vertical lines in the lower three panels show the corresponding results of NSs without a quark core.
An onset density $\sim 6.26\,n_0$ is favored for the small core case based on the joint analysis of $2.14\Msun$ pulsar+GW170817+\nicer, and it is clear that small-core hybrid stars exhibit a maximum mass very close to that of normal hadronic ones.
} 
\vspace{-0.3cm}
\label{fig:multi-posterior of css para for small core case}
\end{figure*}

\section{Conclusions}
\label{sec:sum}

In conclusion, the uncertain maximum mass $M_{\rm TOV}$ 
of NSs has been a crucial problem with ever increasing precision of the observations on NS binaries and pulsars. 
We explicitly include the hadron-quark phase transition in NS EOSs and perform a Bayesian analysis on $M_{\rm TOV}$ of hybrid NSs with a quark core, using the available data from LIGO/Virgo and \nicer.
Two unified hadronic EOSs (the soft QMF and the stiff DD2) are applied in order to test the effect 
due to low-density hadronic matter and 
the model dependence of the results 
for $M_{\rm TOV}$. 
We determine the maximum mass that a hybrid star can reach to be $M_{\rm TOV}=2.36^{+0.49}_{-0.26}\Msun$ ($2.39^{+0.47}_{-0.28}\Msun$), to the 90\% posterior credible level, for soft QMF (stiff DD2), which appears robust with respect to the uncertainties of the hadronic EOS.
This result could have profound implications for the minimum black hole mass, 
as well as help identify the nature of a compact object with 
its mass falling into the possible mass gap. Further discussions along this line have also been provided.

There are several caveats in the present work. 
The major one is our assumption that 
the switching of a hadronic EOS (presently nucleonic EOS) to quark matter EOS is through a 
first-order phase transition. 
The transition may be of second order or even a smooth crossover.
Nevertheless, our framework successfully describes all current laboratory nuclear experiments and astrophysical observations related to NSs. 
The general requirements adopted here (e.g., causality) should also apply to 
any alternative hadron-quark phase transition scenarios or other types of strangeness phase transitions~\citep[see discussions on hyperons, kaons in, e.g.,][]{2020JHEAp..28...19L}.
Therefore, our main conclusions remain valid and useful for identifying compact objects' nature with 
its mass falling into the possible mass gap.
Another assumption that should be taken with caution is about an early onset density 
($\approx2\,n_0$) for the phase transition in the cores of NSs.
We mainly discuss this big-core case because the alternative scenario with a late appearance 
of quarks 
at much higher densities 
($\approx6\,n_0$) results in similar predictions for hybrid stars with respect to normal NSs, where $M_{\rm TOV}$ is primarily determined by the nucleonic interaction instead of properties of the possible phase transitions. 
It is worth mentioning that for normal NSs 
some microscopic calculations including three-body nucleonic interactions~\citep{2019JPhG...46c4001W} incidentally give rise to 
a value of $M_{\rm TOV}$ close to that found for NSs with a quark core in the present work. 
At last, the prior dependence in the Bayesian analysis that can be strongly related to the form of the assumed EOS (for which we chose to use the CSS parameterization) 
has been a well-known restriction, and we plan to examine its effect in more detail in a separate study.

\appendix

\section{Small-core hybrid stars resemble an NS}
\label{sec:A1}

We demonstrate in Fig.~\ref{fig:multi-posterior of css para for small core case} the properties of small-core hybrid stars along with the corresponding EOS parameters within QMF+CSS. 
For such small-core hybrid stars, a strong first-order transition of $\Delta\varepsilon/\varepsilon_{\rm trans}\approx{0.89}$ occurs at $n_{\rm trans}/n_0\approx6.18$, to the 90\% confidence level.
Their maximum masses are very similar to their hadronic counterparts.
In general a hadron-quark phase transition taking place above $n_{\rm trans}\gtrsim 4\,n_0$ is hard to probe~\citep[e.g.,][]{2020ApJ...904..103M}. 
There have been discussions in the literature on probing the sharp phase transition in hybrid stars through tidal deformation measurements, which could be effective when the size of the elastic hadronic region of a hybrid star is large enough~\citep{2020ApJ...895...28P}.

\section{Typical $1.4\Msun$ and massive $2.0\Msun$ hybrid stars}
\label{sec:A2}

\begin{table*}
  \centering
   \vskip-3mm
  \caption{Most probable intervals of various properties for $1.4\Msun$ and $2.0\Msun$ hybrid stars ($90\%$ confidence interval), constrained by four different analyses (explained in Sec.~\ref{sec:priors_constraints}) within the QMF/DD2+CSS framework.
  }
  \setlength{\tabcolsep}{0.8pt}
\renewcommand\arraystretch{1.2}
  \begin{ruledtabular}
  \begin{tabular}{lccccc}      
   Parameters & &  $2.14 \Msun$ pulsar & +GW170817 & +\nicer & +GW170817+\nicer  \\
    \hline \hline 
    \multirow{2}{*}{$R_{1.4}/\rm km$} & QMF  &  $11.66_{-1.70}^{+1.75}$    &  $10.95_{-0.86}^{+1.23}$ & $11.93_{-0.99}^{+1.36}$ &  $11.70_{-0.74}^{+0.85}$ \\
    & DD2  &  $12.36_{-2.02}^{+1.49}$ & $11.10_{-0.83}^{+1.66}$ & $12.58_{-1.60}^{+0.79}$ &  $11.95_{-0.94}^{+1.04}$ \\
    \hline
     \multirow{2}{*}{$\Lambda_{\rm 1.4}$} & QMF   &  $319_{-194}^{+595}$   &  $216_{-85}^{+235}$ & $379_{-188}^{+469}$ &  $312_{-124}^{+254}$ \\
     & DD2  &  $449_{-309}^{+625}$   &  $214_{-79}^{+329}$ & $486_{-305}^{+337}$ &  $333_{-148}^{+298}$ \\
    \hline    
    \multirow{2}{*}{$n^c_{\rm 1.4}/\rm fm^{-3}$} & QMF  &  $0.45_{-0.19}^{+0.20}$ &  $0.52_{-0.15}^{+0.12}$ &  $0.41_{-0.14}^{+0.18}$ & $0.48_{-0.15}^{+0.13}$ \\
    & DD2  &  $0.41_{-0.16}^{+0.23}$ &  $0.52_{-0.15}^{+0.12}$ &  $0.41_{-0.13}^{+0.19}$ & $0.46_{-0.12}^{+0.14}$ \\
    \hline      \hline 
    \multirow{2}{*}{$R_{2.0}/\rm km$} & QMF   &  $11.71_{-1.53}^{+2.47}$    &  $11.16_{-0.92}^{+1.48}$ & $12.11_{-1.37}^{+1.88}$ &  $11.66_{-1.02}^{+1.46}$ \\
    & DD2   &  $12.64_{-2.10}^{+1.84}$ &  $11.23_{-0.86}^{+1.70}$ & $12.65_{-1.73}^{+1.17}$ &  $12.05_{-1.22}^{+1.18}$ \\
    \hline
     \multirow{2}{*}{$\Lambda_{\rm 2.0}$} & QMF   &  $32_{-23}^{+118}$   &  $22_{-12}^{+38}$ & $42_{-30}^{+92}$ &  $29_{-17}^{+54}$ \\
     & DD2  &  $51_{-39}^{+111}$   &  $22_{-12}^{+40}$ & $49_{-35}^{+64}$ &  $36_{-22}^{+44}$  \\
    \hline    
    \multirow{2}{*}{$n^c_{\rm 2.0}/\rm fm^{-3}$} & QMF  &  $0.57_{-0.27}^{+0.32}$ &  $0.65_{-0.21}^{+0.23}$ &  $0.52_{-0.21}^{+0.33}$ & $0.60_{-0.22}^{+0.27}$ \\
    & DD2  &  $0.51_{-0.23}^{+0.32}$ &  $0.65_{-0.20}^{+0.22}$ &  $0.53_{-0.20}^{+0.29}$ & $0.56_{-0.17}^{+0.24}$ \\
  \end{tabular}
  \end{ruledtabular}
  \label{tb:2}
\end{table*}
\begin{figure*}
\centering
\includegraphics[width=7.in]{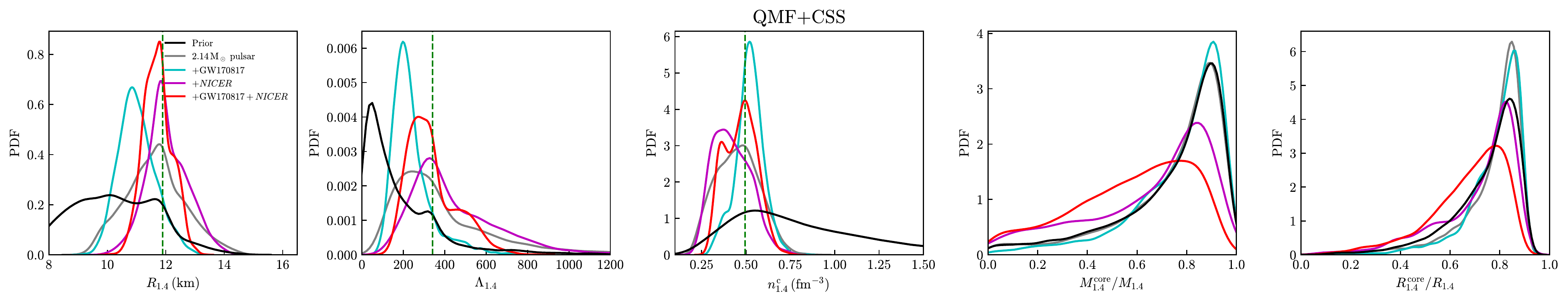}
\includegraphics[width=7.in]{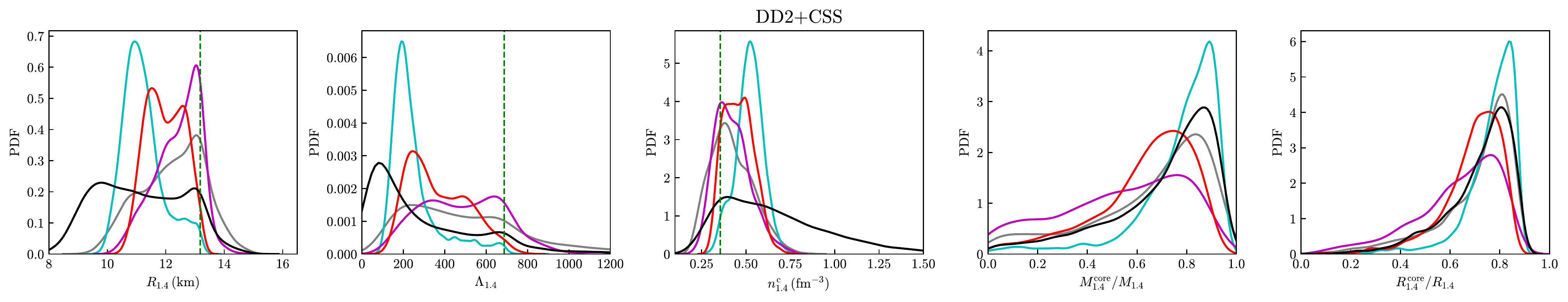}
\caption{Posterior PDFs of various properties for $1.4\Msun$ hybrid stars (90\% confidence level) conditioned on the priors (shown in black lines in each panel) of four different analyses (see details in Sec.~\ref{sec:priors_constraints}).
The green vertical lines in the left six panels show the corresponding results of NSs without a quark core.
} 
\vspace{-0.3cm}
\label{fig:multi-posterior of NS prop for 14}
\end{figure*}

\begin{figure*}
\centering
\includegraphics[width=7.in]{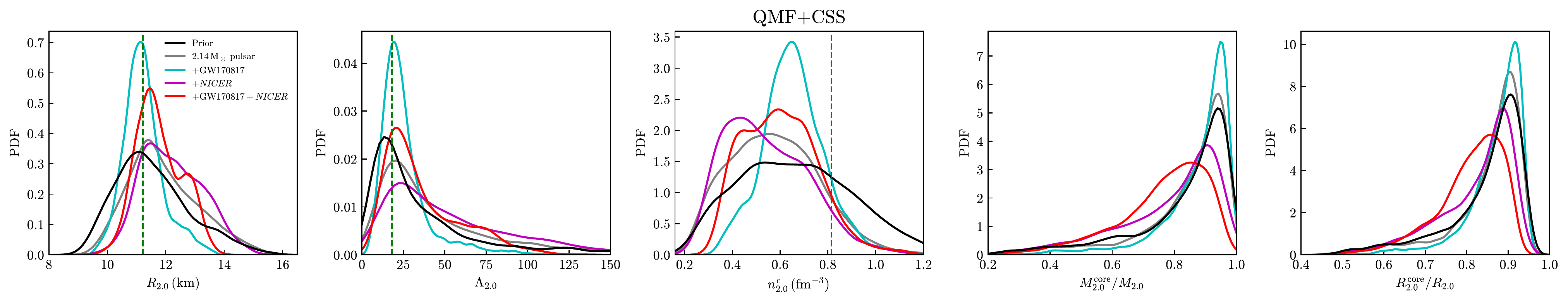}
\includegraphics[width=7.in]{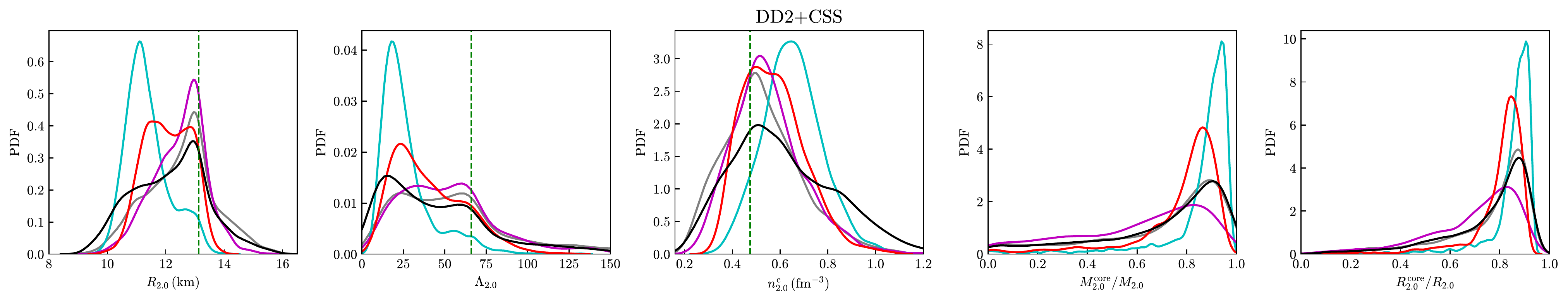}
\caption{Same as Fig.~\ref{fig:multi-posterior of NS prop for 14}, but for $2.0\Msun$ hybrid stars.
} 
\vspace{-0.3cm}
\label{fig:multi-posterior of NS prop for 2}
\end{figure*}

We supplementarily provide various properties of typical hybrid stars with masses $M=1.4\Msun$ and $M=2.0\Msun$ 
in Fig.~\ref{fig:multi-posterior of NS prop for 14}
and Fig.~\ref{fig:multi-posterior of NS prop for 2}, 
and the data of radius, tidal deformability, and central density are summarized in Table \ref{tb:2}.
The results (to the 90\% confidence level) are collected as follows which are conditioned on the joint analysis priors ($2.14\Msun$ pulsar+GW170817+\nicer).

For $1.4\Msun$ hybrid stars, we find that the radius and tidal deformability are $R_{\rm 1.4}=10.96-12.55\,{\rm km}$ and $\Lambda_{\rm 1.4}=188-566$ for the QMF case and $R_{\rm 1.4}=11.01-12.99\,{\rm km}$ and $\Lambda_{\rm 1.4}=185-631$ for the DD2 case,
corresponding to a central density of $n_{\rm 1.4}^{c}=0.48_{-0.15}^{+0.13}~\rm fm^{-3}$ and $0.46_{-0.12}^{+0.14}~\rm fm^{-3}$, respectively.
The probability distribution of the quark matter core in a $1.4\Msun$ hybrid star peaks within a sphere of $R^{\rm core}_{\rm 1.4}/R_{\rm 1.4}\approx0.71$.  
The corresponding mass fraction of quark matter is $\gtrsim62\%$. 
For $2.0\Msun$ hybrid stars, it is found that $R_{\rm 2.0}=10.64-13.12\,{\rm km}$ and $\Lambda_{\rm 2.0}=12-83$ for the QMF case, and $R_{\rm 2.0}=10.83-13.23\,{\rm km}$ and $\Lambda_{\rm 2.0}=14-80$ for the DD2 case, corresponding to a central density of $n_{\rm 2.0}^{c}=0.60_{-0.22}^{+0.27}~\rm fm^{-3}$ and $0.56_{-0.17}^{+0.24}~\rm fm^{-3}$, respectively.
The probability distribution of quark matter core in a $2.0\Msun$ hybrid star peaks within a sphere of $R^{\rm core}_{\rm 2.0}/R_{\rm 2.0}\approx0.83$.
The corresponding mass fraction of quark matter is increased to $\gtrsim79\%$. 

\section{Adding the GW190425 tidal deformability data}
\label{sec:A3}

\begin{figure*}
\label{fig:tran_para_add}
\centering
\includegraphics[width=6.5in]{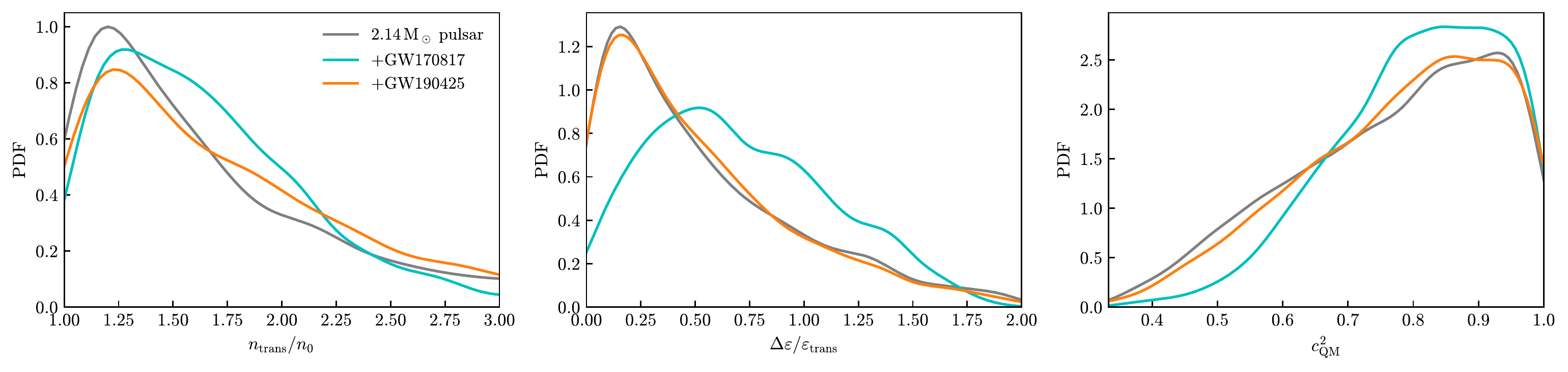}
\includegraphics[width=6.5in]{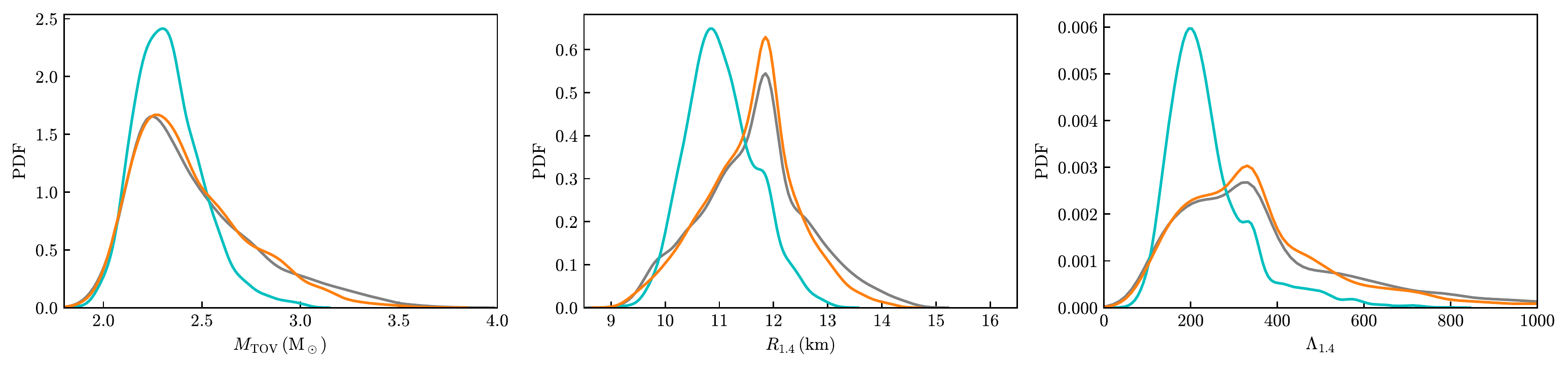}
\caption{Posterior PDFs of three CSS parameters (upper panels) and the hybrid star maximum mass (lower panels), the radius and tidal deformability of a $1.4\Msun$ hybrid star, based on the QMF+CSS modeling of dense matter EOS for three different analyses.
} 
\vspace{-0.3cm}
\label{fig:tran para with additional data}
\end{figure*}

The recent reported GW190425's component masses range from $1.46$ to $1.87\Msun$ ($1.12–2.52\Msun$ for high spin priors)~\citep{2020ApJ...892L...3A} are compatible with existing constraints on the NS maximum mass ($\sim2.10-2.85\Msun$), suggesting that the event system might be an NS binary.
We therefore in this Appendix add the GW190425 event into the analysis and discuss its effects on the parameter space of the theoretical model and the hybrid star properties. 

In Fig.~\ref{fig:tran para with additional data} we present the posterior PDFs of the CSS parameters $(n_{\rm trans}/n_0,\Delta\varepsilon/\varepsilon_{\rm trans},c_{\rm QM}^2)$ for different analyses 
as well as those of the hybrid star properties i.e., the maximum mass, the radius and tidal deformability of a $1.4 \Msun$ star. 
The GW190425 results are found, in most cases, very similar to that based on the $2.14 \Msun$ pulsar constraint, and the additional GW190425 data provides little improvement to the results (mainly due to the intrinsically much smaller tidal deformability and its low signal-to-noise ratio).

\acknowledgments
We are thankful to Tong Liu, Wen-Jie Xie for helpful discussions. 
The work is supported by National SKA Program of China (No.~2020SKA0120300), the National Natural Science Foundation of China (Grant No.~11873040) and the Youth Innovation Fund of Xiamen (No.~3502Z20206061). 
S.H. acknowledges support from the National Science Foundation, Grant PHY-1630782, and the Heising-Simons Foundation, Grant 2017-228.
This research has made use of data and software obtained from the Gravitational Wave Open Science Center $https://www.gw-openscience.org$, a service of LIGO Laboratory, the LIGO Scientific Collaboration and the Virgo Collaboration. LIGO is funded by the U.S. National Science Foundation. Virgo is funded by the French Centre National de la Recherche Scientifique (CNRS), the Italian Istituto Nazionale di Fisica Nucleare (INFN) and the Dutch Nikhef, with contributions by Polish and Hungarian institutes.

\software{Bilby \citep[version 0.5.5, ascl:1901.011, \url{https://git.ligo.org/lscsoft/bilby/}]{2019ascl.soft01011A}, PyMultiNest \citep[version 2.6, ascl:1606.005, \url{https://github.com/JohannesBuchner/PyMultiNest}]{2016ascl.soft06005B}.}


\end{document}